\documentclass[9pt,twocolumn,twoside]{opticajnl}
\journal{opticajournal} 
\setboolean{shortarticle}{true}
\usepackage{lineno}
\usepackage{float}
\usepackage{siunitx}
\usepackage{stfloats}

\usepackage{soul}
\usepackage{color}

\setlist[itemize]{noitemsep}

\title{Average-power scalability of multi-cycle terahertz sources based on periodically poled lithium niobate stacks}
\author[1,2,\dag,*]{Patrick J. Dalton}
\author[3,\dag,*]{Robin L\"{o}scher}
\author[3]{Tim Vogel}
\author[1,2]{Robert B. Appleby}
\author[1,4]{Graeme Burt}
\author[1,5]{Steven P. Jamison}
\author[1,2]{Morgan T. Hibberd}
\author[1,2]{Darren M. Graham}
\author[3,6]{Clara J. Saraceno}

\affil[1]{The Cockcroft Institute, Sci-Tech Daresbury, Keckwick Lane, Warrington WA4 4AD, United Kingdom}
\affil[2]{Department of Physics and Astronomy \& Photon Science Institute, The University of Manchester, Manchester M13 9PL, United Kingdom}
\affil[3]{Photonics and Ultrafast Laser Science, Ruhr-Universit\"{a}t Bochum, Bochum 44801, Germany}
\affil[4]{School of Engineering, Lancaster University, Bailrigg, Lancaster LA1 4YW, United Kingdom}
\affil[5]{Department of Physics, Lancaster University, Bailrigg, Lancaster LA1 4YB, United Kingdom}
\affil[6]{Research Center Chemical Science and Sustainability, University Alliance Ruhr, Bochum 44801, Germany}
\affil[*]{patrick.dalton@manchester.ac.uk, robin.loescher@ruhr-uni-bochum.de}
\affil[$\dagger$]{These authors contributed equally to this work.}

\begin{abstract}
We demonstrate that narrowband multi-cycle terahertz (MC-THz) sources based on periodically-poled lithium niobate (PPLN) wafer stacks can be driven by high repetition-rate, high energy femtosecond ytterbium-doped lasers. Operating at 10-kHz repetition rate with up to 104\,W of pump power on a 10-wafer stack, we measure 26.4\,mW of THz average power for a  narrowband multi-cycle source. We identify and quantify strong lensing effects causing dramatic beam focusing in 47 wafer stacks which act as a primary limitation in the current configuration, and present mitigation strategies for future scaling. This first study of high average power narrowband multi-cycle THz sources offers a path forward to Watt-level high repetition rate sources using thin lithium niobate plates.
\end{abstract}

\setboolean{displaycopyright}{false} % Do not include copyright or licensing information in submission.

\begin{document}

\maketitle

Narrowband multi-cycle THz (MC-THz) sources are of interest for a large number of applications including the selective excitation of resonant modes in materials \cite{dienst_optical_2013}, harmonics generation \cite{hafez_extremely_2018}, and compact particle accelerators \cite{Hibberd2020}. Growing interest has driven many recent advances \cite{fulop_laserdriven_2020}, with efforts focusing mainly on pulse energy scaling \cite{matlis_scaling_2024}. 

Another important parameter with relevance to the applications of these sources is their repetition rate. Higher repetition rates enable faster THz transient acquisition, rapid statistics, and improved experimental efficiency, particularly in electron acceleration \cite{rovige_optimization_2021}. However, little effort has been made to make use of high average power ytterbium (Yb) drivers to increase the repetition rate of these narrowband sources beyond 1\,kHz. Most efforts so far to increase average power of THz sources have concentrated on single-cycle broadband THz sources \cite{loscher_laser-driven_2025}.

The most commonly used approach for generating MC-THz pulses is nonlinear optical down-conversion of femtosecond lasers in large aperture, periodically-poled lithium niobate (PPLN). Two types of approaches have been studied so far to reach high energies: traditional voltage-poled bulk material \cite{lhuillier_generation_2007}, and stacks of large aperture sub-millimeter thick wafers in alternating orientations \cite{Lemery2020}. Stacked lithium niobate (LN) plates are particularly interesting due to their tunability by the number of wafers stacked, \textit{i.e.} the pump beam interaction length, to control the number of cycles and tailor the generation process for the desired applications \cite{Mosley2023}. Furthermore, much larger aperture wafers are available than bulk crystals, allowing straightforward energy scaling with large apertures. A comparison of both techniques has been recently shown in \cite{matlis_scaling_2024} using PPLN and KTP, highlighting the challenges related to very high energy excitation in LN, such as multi-photon absorption, infrared induced absorption and others. At high average powers, these challenges are known to become more severe due to cumulative effects and additional thermal management difficulties. For example, the thermal load at elevated repetition rates will result in changes of refractive index and absorption of the material in both the optical and THz beams \cite{Wu2015, benabdelghani_three-photon_2024}, which leads to effects such as thermal lensing or depolarization. Although the thermo-optic and thermal conductivity coefficients in LN are well known, the impact of these effects for high average power THz sources have only recently started to be studied in the context of the titled-pulse front technique for broadband single cycle pulses \cite{loscher_laser-driven_2025}, and there are no reports of increasing the repetition rate of PPLN-based MC-THz sources beyond 1\,kHz to explore the influence of thermal effects.  

In this letter, we generate narrowband MC-THz radiation in a PPLN wafer stack (PPLN-WS) driven by a Yb-based thin-disk amplifier laser system operating at 1030-nm wavelength, 10-kHz repetition rate, and 700-fs pulse width. By pumping the PPLN-WS with 104\,W of average power, we measure 26.4\,mW of THz average power. We identify a combination of photorefractive-, thermal-, and Kerr-lensing in the PPLN as the main limitation for our approach, warranting further studies on the interplay of pump peak and average power for future Watt-class MC-THz sources. Our work presents first steps towards improved measurement statistics and rapid experiments, for example in laser-driven particle acceleration at higher repetition rates.

\begin{figure}
    \centering
    \includegraphics[width=.9\linewidth]{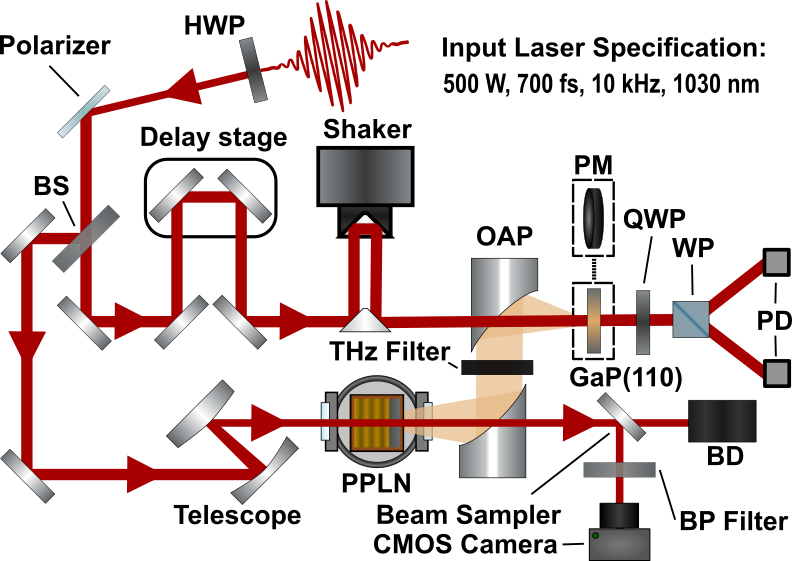}
    \caption{Schematic diagram of the THz generation and detection setup. HWP = half-wave plate, BS = beam splitter, OAP = off-axis parabolic mirror, QWP = quarter-wave plate, WP = Wollaston prism, PD = balanced photodetector, BP Filter = bandpass filter, PM = power meter, BD = beam dump. }
    \label{fig:Figure1}
\end{figure}

\begin{figure*}[!t]
    \centering
    \includegraphics[width=.9\textwidth]{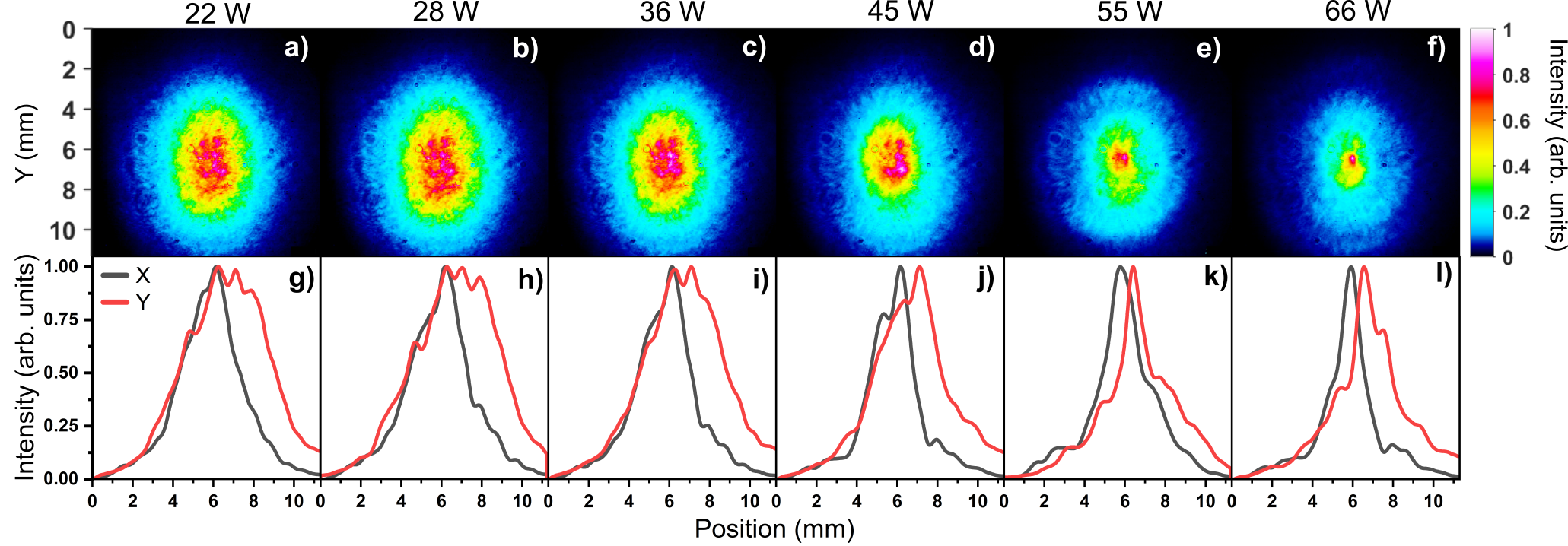}
    \caption{Unconverted pump beam profiles measured after transmission through a PPLN stack (47 wafers) for average pump powers between a) 22\,W and f) 66\,W. Corresponding \textit{x}- and \textit{y}-line profiles in subfigures g) to l).}
    \label{fig:Figure2}
\end{figure*}

\textbf{Experimental setup.} 
In this experiment we used 5-mol.\% magnesium oxide-doped congruent \textit{x}-cut LN wafers with a diameter of 50.8\,mm and a nominal thickness of 135\,µm with an anti-reflection coating for the pump wavelength. The wafer thickness corresponded to a poling period of 270\,µm, and a THz output frequency around 0.4\,THz in the forward propagation direction. A 500-µm thick \textit{z}-cut quartz wafer was added to reduce Fresnel losses at the LN-vacuum interface for optimized out-coupling of the THz radiation.

The driving laser is a thin-disk regenerative amplifier system (DIRA 500-10, TRUMPF Scientific Lasers) and the experimental setup is shown in Fig. \ref{fig:Figure1}. The pump average power applied to the PPLN was controlled by a half-waveplate and a thin-film polarizer with calibrated average power in front of the PPLN-WS. The output was split into a pump and gate pulses for electro-optic sampling (EOS) by an output coupler with 0.2\% transmission. The pump beam area was reduced to 6.04\,$\times$ 7.51\,mm (1/$\mathrm{e}^2$) using a Galilean mirror-based telescope.  Two 90$^{\circ}$ off-axis parabolic mirrors (OAP) with focal lengths of 101.6\,mm, and central 3-mm diameter holes, were used to collect the generated THz radiation and focus it onto either a (110)-cut 1-mm thick gallium phosphide (GaP) crystal for EOS, or a thermopile sensor (3A-P-THz, Ophir) for THz average power measurements. The hole in the first parabolic mirror allowed the unconverted pump power to be directed into a beam dump, while the hole in the second parabolic mirror allowed for EOS with a collinear probe beam. A THz transmitting filter was placed in the collimated THz beam path between the two OAPs to block residual 1030-nm pump light and any 515-nm second-harmonic generated in the PPLN-WS. The filter was made from a combination of polytetrafluoroethylene (PTFE) and high-density polyethylene (HDPE) plates. Values for the THz average power within this letter are corrected for reflective losses at the filters and transmission losses at the holes of the OAPs. 

The EOS measurements employed a sinusoidally moving delay line to provide a \~40-ps delay range for the gate pulse and an open-source algorithm was used for data processing \cite{vogel_advanced_2024}. An additional motorized delay stage allowed the acquisition of THz waveforms over longer time delays by stitching individual 40-ps slices. The stitching algorithm utilized a correlation function for optimized delay shifting. The unconverted pump beam profile exiting the PPLN-WS was measured by removing the first OAP and reflecting 0.05\% with a beam sampler through a 1030-nm bandpass filter (10-nm FWHM) and onto a CMOS beam profiler, positioned 15\,cm from the exit of the PPLN-WS. 

\textbf{Results and discussion.} We first examined lensing effects in a 47-wafer PPLN stack (~6.345 mm thick) under high average power. Figures \ref{fig:Figure2}a)-f) show the unconverted pump beam profiles after transmission through the 47-wafer PPLN stack as the average pump power is varied from 22\,W to 66\,W, with the corresponding \textit{x}- and \textit{y}-line profiles through the peak intensity presented in Figs. \ref{fig:Figure2}g)-l). The thick stack produces a dramatic focusing effect that intensifies with increasing pump power. As the pump power increases from 22\,W to 66\,W, the FWHM decreases substantially from 3.04\,mm to 1.40\,mm in the \textit{x}-direction and from 4.87\,mm to 1.99\,mm in the \textit{y}-direction.

\begin{figure}[!ht]
    \centering
    \includegraphics[width=\linewidth]{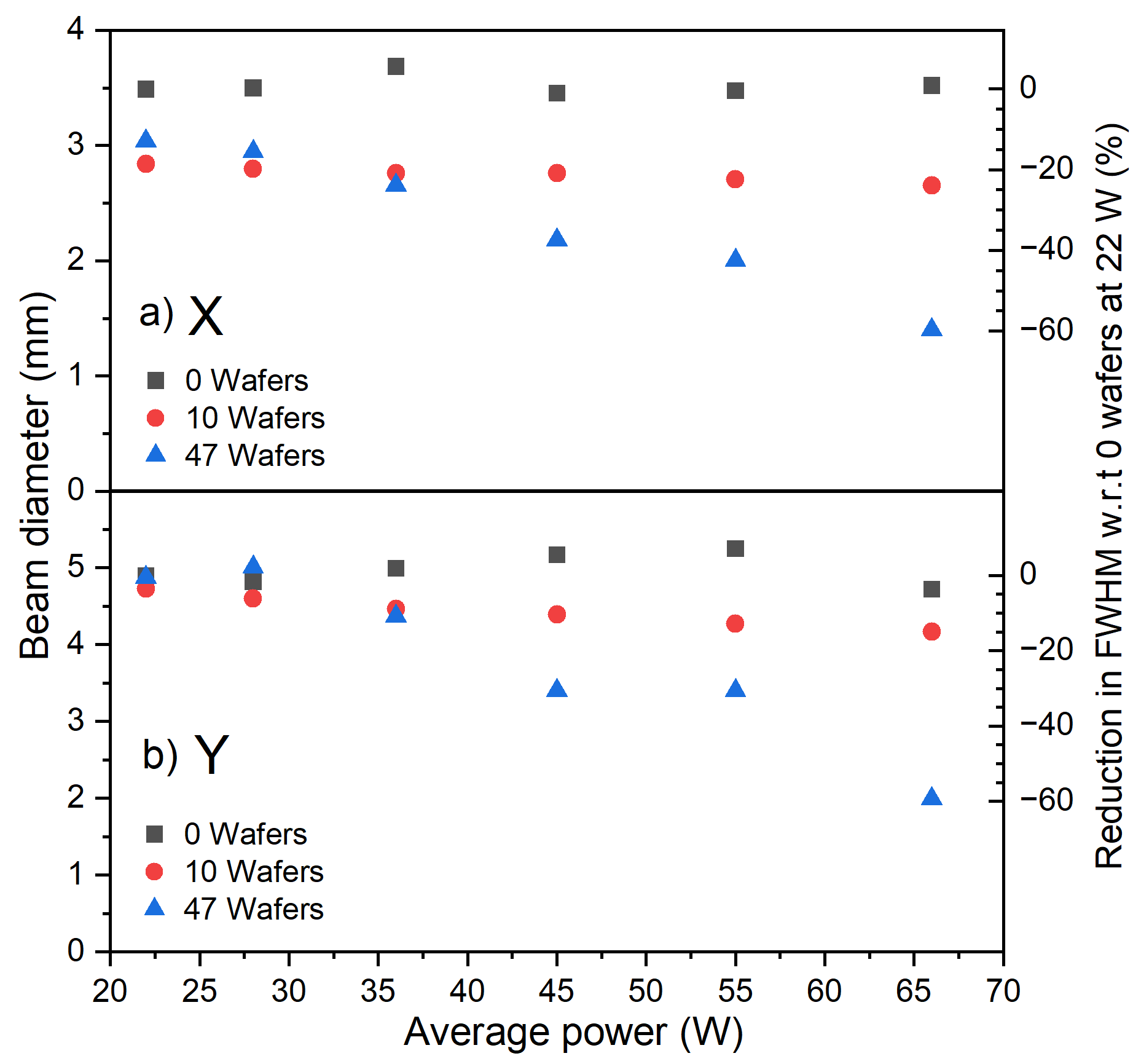}
    \caption{FWHM beam diameter of the unconverted pump from a) \textit{x}- and b) \textit{y}-line profiles after transmission through stacks of 0, 10 and 47 PPLN wafers.}
    \label{fig:Figure3}
\end{figure}

To compare the lensing behavior across different stack thicknesses, we measured unconverted pump beam profiles for 0, 10, and 47-wafer configurations. Figure \ref{fig:Figure3} summarizes the FWHM beam diameters extracted from the \textit{x}- and \textit{y}-line profiles, revealing the progression of those lensing effects. All configurations show similar beam dimensions at low powers, indicating minimal lensing. However, above 45\,W, a clear dependence on stack thickness emerges, with thicker stacks producing stronger focusing. At 45\,W, the \textit{x}-direction FWHM decreases progressively from 3.45\,mm (no wafers) to 2.76\,mm (10 wafers) to 2.18\,mm (47 wafers), while the \textit{y}-direction shows a similar trend: 5.17\,mm, 4.39\,mm, and 3.40\,mm, respectively. This lensing effect intensifies further at 66\,W, where the 47-wafer stack produces dramatic beam compression with a FWHM of 1.40\,mm (\textit{x}) and 1.99\,mm (\textit{y}), compared to 3.52\,mm and 4.72\,mm respectively for the reference case without wafers.

The consequence of these strong lensing effects is a reduction of the pump beam area and a dramatic increase in the pump beam fluence which ultimately results in damage to optical components after the PPLN-WS and the generation of an air-plasma. This then limits the options available for separating out the residual pump beam from the generated THz radiation, and therefore presents a barrier to increasing the average pump power up to the maximum 500\,W available from the laser system in the current setup. We also note that the change in beam size increases the intensity during propagation trough the PPLN-WS, therefore making the modeling of THz generation process more involved. In future, 3d+1 pulse propagation simulations should be used to capture all propagation effects through the plates. 

To better understand the relative contributions of different lensing mechanisms — Kerr lensing, photorefractive effects, and thermal lensing — we performed a simple calculation based on the optical Kerr effect. Using a nonlinear refractive index coefficient of $n_2 = \SI{9.33e-20}{\centi\meter\squared\per\watt}$ for LN \cite{DeSalvo:1996}, we modeled the PPLN-WS and the surrounding windows individually as single bulk structures divided into 1000 slices with corresponding free-space propagation distances. Each slice was treated as a thin lens with a focal length determined by the Kerr-lens equation, and the effect of the consecutive lenses was calculated using ABCD-matrix propagation. Focusing on the \textit{x}-axis with more severe Kerr lensing, the beam sizes at the camera position in our model decreased by relative 2.81\%, 3.07\%, and 3.11\% for 0, 10, and 47 wafer configurations, respectively. Thus, Kerr-lensing alone cannot account for the strong focusing in our experiment (one order of magnitude higher, see Fig. \ref{fig:Figure3}), indicating that photorefractive or thermal lensing (or a combination of both) inside the PPLN-WS play the dominant role in our high-average-power regime \cite{benabdelghani_three-photon_2024, henderson_intra-cavity_2006, loscher_laser-driven_2025}. However, the wafers used were MgO-doped which is known to suppress photorefractive effects, suggesting thermal lensing may be the dominant effect \cite{10.1063/1.1318721}.

\begin{figure}[!h]
    \centering
    \includegraphics[width=.9\linewidth]{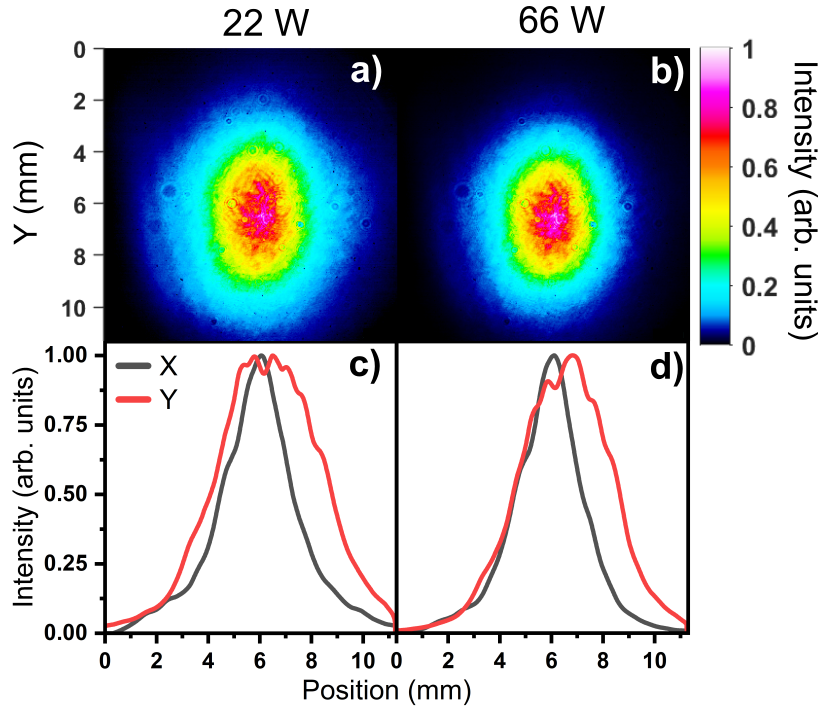}
    \caption{Unconverted pump beam profiles measured after transmission through a PPLN stack (10 wafers) for average pump powers of a) 22\,W and b) 66\,W. Extracted \textit{x}- and \textit{y}-line profiles through the profile peak intensity for average pump powers of c) 22\,W and d) 66\,W.}
    \label{fig:Figure4}
\end{figure}

Within the limits of our setup imposed by this strong thermal lensing, we therefore restricted ourselves to a stack with 10 wafers (corresponding to a thickness of 1.350\,mm). This configuration provided a manageable balance between THz generation efficiency and optical damage thresholds. Figures \ref{fig:Figure4}a) and b) show the unconverted pump beam profiles after transmission through the 10-wafer PPLN stack at average pump powers of 22\,W and 66\,W, respectively, with the corresponding \textit{x}- and \textit{y}-line profiles through the peak intensity shown in Figs. \ref{fig:Figure4}c) and d). At the lower pump power of 22\,W, the beam exhibits an elliptical profile with full-width at half-maximum (FWHM) values of 2.84\,mm in \textit{x} and 4.73\,mm in \textit{y}. As the average pump power increases to 66\,W, a noticeable beam focusing effect becomes apparent, with the FWHM decreasing to 2.65\,mm in \textit{x} and 4.17\,mm in \textit{y}, indicating a reduction in the overall beam area. However, this lensing effect is significantly more manageable compared to the 47-wafer configuration \ref{fig:Figure3}. In this 10-wafer configuration, we were able to utilize a pump laser power of up to 104\,W (at 10\,kHz) to generate an average THz power of 26.4\,mW. The measurements were performed using a thermopile sensor (3A-P-THz, Ophir) calibrated by the manufacturer for the spectral range from 0.1 to 30\,THz. We corrected the measured average power for the transport loss through the OAPs and the transmission lost through the THz filter. However, the absorptive glass of the thermopile sensor is highly transmissive at frequencies below 700\,GHz \cite{Steiger:13}, rendering the calibration questionable in this range. Consultation with the Physikalisch-Technische Bundesanstalt (PTB), the German national metrology institute, confirmed that no facility worldwide currently provides calibrations for THz detectors at these low frequencies. For future experiments, the use of spectrally flat, calibrated pyroelectric THz detectors is advised for accurate power metrology of the generated THz radiation. However, in previous work of our group we confirmed the performance of the detector used in this study with a spectrally flat, calibrated detector \cite{khalili_high-power_2025}. Within this uncertainty, the current value is the highest so far demonstrated with MC-THz sources.

\begin{figure}[h]
    \centering
    \includegraphics[width=.9\linewidth]{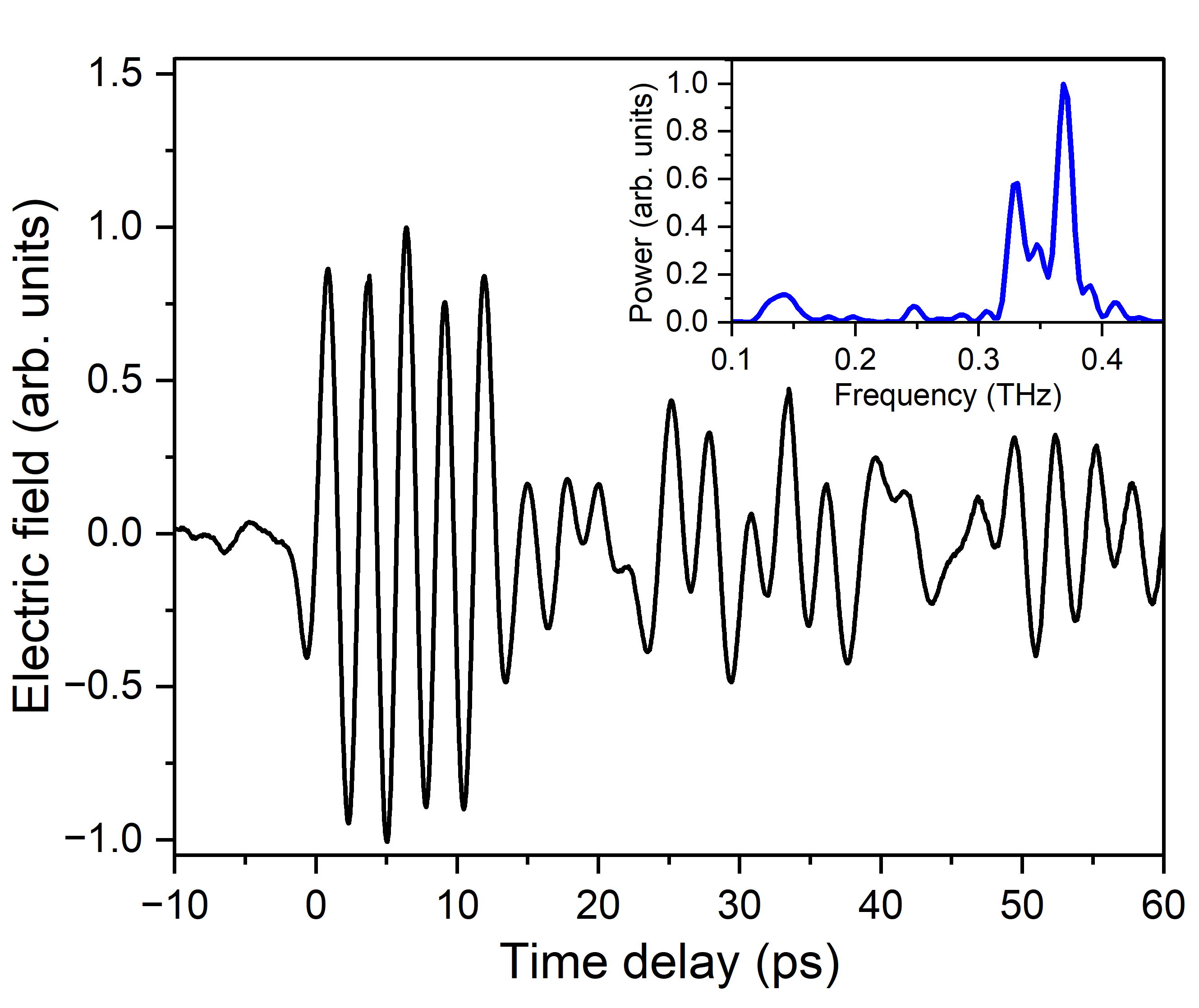}
    \caption{THz waveform and corresponding power spectrum recorded for a PPLN stack (10 wafers) when pumped with an average power of 28.4\,W at 10\,kHz of repetition rate.}
    \label{fig:Figure5}
\end{figure}

Figure \ref{fig:Figure5} shows the waveform acquired for the 10-wafer PPLN stack when pumped with an average power of 28.4\,W at 10 kHz of repetition rate. The five-cycle THz pulse is shown between 2.3 and 14.3\,ps. Additional features in the waveform correspond to various reflection artifacts: pump beam reflections within the PPLN stack (12.6\,ps), THz reflections from the 1-mm thick GaP detection crystal (15.7\,ps), THz reflections within the PPLN stack (39.8\,ps), and combined pump and THz reflections within the stack (52.4\,ps). The structure of the PPLN-WS forms an optical cavity with multiple internal reflections. Considering the refractive index of LN at our pump wavelength \cite{Zelmon:97}, we can attribute the artifacts to originate from various depths inside our PPLN stack. The inset of Fig. \ref{fig:Figure5} shows the power spectrum, revealing a dominant 0.37-THz peak from forward-propagating THz generation and a secondary 0.14-THz peak from backward-generated THz that reflects at the entrance interface and propagates forward. This is in agreement with previous measurements at significantly lower repetition rates \cite{Dalton2024}.

\textbf{Conclusion and outlook.}
We present a first study of average-power scaling for narrowband MC-THz sources based on PPLN-WS using a high repetition-rate ytterbium-based thin-disk amplifier system. We identify and quantify lensing effects as the main limiting factor in our experimental setup. By reducing the number of LN plates to ten while sacrificing efficiency, we mitigate these effects and measure a THz average power of 26.4\,mW at 104 W of driving power and 10 kHz repetition rate.

Our results present guidelines for future scaling to a Watt-level MC-THz source provided these thermal effects can be mitigated. A 0.17\% conversion efficiency, as previously obtained at high energy \cite{Mosley2023} , together with the full pulse energy available from the system (50\,mJ at 10\,kHz) would lead to 0.85\,W of average power. Since the PPLN wafers were not damaged, the beam focusing could be managed within a vacuum environment. Additionally, given that simulations and experiments isolate thermal lensing as the dominant effect, smaller wafer diameters that still support the pulse energies available at these average powers would improve the heat conductance, mitigating the severe focusing effects, and enabling utilization of the full 500-W pump power available from the laser system. Aside from these engineering efforts, further experimental and theoretical studies are required to understand and overcome the observed lensing effects, enabling the power-scaling of MC-THz sources based on PPLN-WS.

\begin{backmatter}
\bmsection{Funding} United Kingdom Science and Technology Facilities Council (STFC) [Grant No. ST/V001612/1]. STFC studentship to Patrick J. Dalton (Project Ref. 2905174). Deutsche Forschungsgemeinschaft (EXC-2033 (390677874-RESOLV). Ministerium für Kultur und Wissenschaft des Landes Nordrhein-Westfalen (“terahertz.NRW”, program “Netzwerke 2021”).
\bmsection{Acknowledgment} We acknowledge the fruitful discussions with Andreas Steiger and Benjamin R\"{o}ben from the Physikalisch-Technische Bundesanstalt, Germany, regarding accurate THz power metrology.

\bmsection{Disclosures} The authors declare no conflicts of interest.

\bmsection{Data availability} Data underlying the results presented in this paper are available in \cite{Dalton2025Zenodo}.
\end{backmatter}

% Bibliography
\bibliography{UoMRuBPaper}

\newpage

% Full length Bibliography
%\bibliographyfullrefs{UoMRuBPaper}

\end{document}